\documentclass[%
 aip,
 amsmath,amssymb,
preprint,%
]{revtex4-1}

\usepackage{graphicx}% Include figure files
\usepackage{dcolumn}% Align table columns on decimal point
\usepackage{bm}% bold math
\usepackage[utf8]{inputenc}
\usepackage[T1]{fontenc}
\usepackage{mathptmx}
\usepackage{etoolbox}
\usepackage{color}
\newcommand{\rev}[1]{{\color{black}#1}}
\newcommand{\revred}[1]{{\color{black}#1}}
\makeatletter
\def\@email#1#2{%
 \endgroup
 \patchcmd{\titleblock@produce}
  {\frontmatter@RRAPformat}
  {\frontmatter@RRAPformat{\produce@RRAP{*#1\href{mailto:#2}{#2}}}\frontmatter@RRAPformat}
  {}{}
}%
\makeatother
\begin{document}
\newcommand{\stefania}[1]{\textcolor{blue}{#1}}
\newcommand{\comment}[1]{\textcolor{green}{#1}}
\newcommand{\Ashahar}[1]{\textcolor{red}{#1}}
%\preprint{AIP/123-QED}
\title{Dynamical control of light-matter interaction through coherent multipolar scattering in broken symmetry metasurfaces}
%\title{Zero-Phase Resonance Engineering in Broken-Symmetry Metasurfaces for Enhanced Emission in Telecom-C-band }
% Force line breaks with \\
\author{Mohammed Ashahar Ahamad}
\email{Corresponding author: ashaharamu2014@gmail.com}

\affiliation{
Department of Physics, Aligarh Muslim University,
Aligarh, Uttar Pradesh 202002, India.
}

\author{Faraz Ahmed Inam}
%\email{faraz.inam.phy@amu.ac.in}
\affiliation{
Department of Physics, Aligarh Muslim University,
Aligarh, Uttar Pradesh 202002, India.
}

\date{\today}

 \begin{abstract}
 Achieving control over spontaneous emission by tailoring light-matter interactions is a key objective in quantum nanophotonics. Metasurfaces composed of high-refractive-index resonators like silicon (Si) provide a low-loss platform that supports a variety of strong electric and magnetic multipolar resonances, offering new opportunities to tailor the local density of optical states (LDOS). \rev{This work employs phase-resolved multipolar analysis to investigate spontaneous-emission control in symmetry-broken dielectric metasurfaces composed of Si cuboid and disk resonators. Controlled in-plane geometrical asymmetry enables hybridization between magnetic dipole (MD) and magnetic quadrupole (MQ) modes, whose coherent interaction modifies the local density of optical states (LDOS). Symmetry-broken dielectric metasurfaces are closely related to quasi-bound states in the continuum (quasi-BICs) and Fano-resonant systems, where geometrical perturbations promote coupling between weakly radiative and bright resonances. From this perspective, the observed MD--MQ hybridized states may be understood as a near-field and multipolar manifestation of symmetry-broken quasi-BIC physics. The resulting coherent magnetic multipolar interaction produces strong near-field localization and substantial spontaneous-emission enhancement for embedded emitters. For Erbium ions (Er$^{3+}$) in broken-symmetry Si metasurfaces operating near 1.54~$\mu$m, the analysis reveals pronounced enhancement associated with resonant magnetic multipolar coupling. These results establish magnetic multipolar interference as a physically transparent mechanism for emission control in low-loss dielectric metasurfaces and provide useful design guidelines for integrated quantum photonic devices.}\\
\end{abstract}

\maketitle
\section{\label{sec:level1} \textit{Introduction} }
Over the past decades, rigorous research has been conducted to control light-matter interactions at the nanoscale, a central objective in the development of efficient quantum photonic devices\cite{gonzalez2024light,koo2024dynamical,Lodahl2015}.A key aspect of this control concerns the spontaneous emission dynamics of quantum emitters, which are not solely determined by their intrinsic properties but are strongly influenced by the surrounding photonic environment through the local density of optical states (LDOS) as described by Fermi’s golden rule
\cite{Purcell1946,Novotny2011,Barnes2020,Pelton2015}. Engineering this environment provides a powerful route to tailor emission strength, directionality, and radiative efficiency at the nanoscale 
\cite{Koenderink2006,Koenderink2017,bohn2018active, Pelton2015,Zahedian2023,ahamad2024electromagnetic}.
A promising route to achieve such control is offered by optical metasurfaces, two-dimensional arrays of subwavelength nanostructures that enable precise manipulation of electromagnetic waves and light-matter interactions \cite{wang2018metasurfaces,Solntsev2021,LiSinghSievenpiper,app9132727,Glybovski2016}. Metasurfaces composed of dielectric resonators, such as silicon (Si), support a variety of electric and magnetic multipolar resonances while avoiding the strong dissipative losses, unlike plasmonic systems \cite{decker2016resonant,Koshelev2020,Ahamad2023,jiang2025comparative,kuznetsov2016optically}. These features make dielectric metasurfaces well suited for controlling emitter-field interactions and spontaneous emission processes at the nanoscale \cite{liu2018light,wang2018metasurfaces}.
The ability of dielectric metasurfaces to tailor emission dynamics is strongly governed by the symmetry of the metasurface unit cell, which determines the accessibility and radiative nature of the supported multipolar modes \cite{liu2018light}. In highly symmetric metasurfaces, only multipolar modes that satisfy symmetry-imposed selection rules, commonly referred to as bright modes, can efficiently couple to free-space radiation \cite{campione2016broken}. Deliberate breaking of structural symmetry lifts these constraints, enabling hybridization between bright and dark multipolar modes. From a current-distribution perspective, symmetry breaking introduces finite spatial overlap between otherwise orthogonal multipolar channels, allowing nominally dark modes to acquire radiative character through coupling with electric dipole–like modes \cite{limonov2017,koshelev2018assymetric}. The resulting interference between these modes often manifests as sharp Fano-type resonances in the optical response, fundamentally reshaping the metasurface’s modal landscape and opening new pathways for controlling light–matter interactions.
\rev{Such symmetry-broken dielectric metasurfaces are closely related to symmetry-protected bound states in the continuum (BICs) and quasi-BICs, where weakly radiative modes become externally accessible only after a controlled perturbation breaks the protecting symmetry. In perfectly symmetric dielectric metasurfaces, certain resonances remain decoupled from the radiation continuum because of symmetry incompatibility, whereas a slight in-plane perturbation opens a radiative channel and converts the ideal BIC into a high-$Q$ quasi-BIC accompanied by a Fano-type spectral response and strong near-field confinement \cite{li2019Bic,Shi2022, Hong2025single}. This general physical picture is consistent with recent studies of all-dielectric metasurfaces in which symmetry breaking, Mie-resonant mode engineering, and quasi-BIC formation are used to tailor sharp optical responses, enhanced local fields, and polarization-dependent scattering \cite{kuznetsov2016optically,li2019Bic,Shi2022, Hong2025single,koshelev2018assymetric}.}
\rev{ We note that the broken cuboid and disk metasurfaces considered here are not designed as chiral metasurfaces, and the present work does not rely on circular dichroism or optical activity as the operating principle. Nevertheless, the broader literature shows that symmetry reduction in dielectric metasurfaces can also be used to generate planar chiral responses and polarization-selective quasi-BIC resonances when the geometry lacks the appropriate mirror symmetries \cite{Shi2022,Hong2025single,xie2020phase,sarma2015rotating}. In the present case, the relevant consequence of symmetry breaking is instead the activation and hybridization of magnetic multipolar channels that modify the LDOS experienced by an embedded emitter.}
\rev{In this broader context, the present work does not introduce a new multipolar formalism; rather, it employs phase-resolved multipolar analysis as an interpretive framework to elucidate how symmetry-induced MD--MQ hybridization modifies the electromagnetic environment experienced by embedded emitters.}
\rev{Despite significant progress in understanding broken-symmetry and quasi-BIC metasurfaces, most existing studies primarily emphasize far-field observables such as reflectance, transmittance, chirality, and scattering spectra \cite{malek2026broken,rozman2024broken,campione2016broken,Shi2022,Hong2025single}. By comparison, the near-field multipolar interactions that directly govern the LDOS experienced by a quantum emitter are often inferred indirectly. In particular, a detailed emitter-centered interpretation linking symmetry breaking, multipolar phase relations, and spontaneous-emission modification remains comparatively less explored.}

\rev{In this work, we employ multipolar decomposition and phase-resolved analysis to investigate symmetry-broken dielectric metasurfaces with a particular focus on spontaneous-emission enhancement.} Using silicon cuboid and disk-based metasurfaces as representative platforms, we begin from fully symmetric configurations and introduce controlled geometrical asymmetry as a continuously tunable parameter. This approach enables systematic tracking of multipolar mode hybridization, phase evolution, and emitter-mode coupling from the symmetric limit to the broken-symmetry regime. Through wavelength-resolved multipolar decomposition and phase analysis, we demonstrate that symmetry breaking enables constructive interference between magnetic dipole (MD) and magnetic quadrupole modes (MQ), driving them toward an in-phase condition at the erbium (Er$^{3+}$) emission wavelength of 1540~nm. This coherent multipolar scattering enhances the local photonic environment experienced by the emitter, leading to a pronounced increase in the spontaneous emission rate.

Our results establish symmetry-controlled multipolar interference as a robust and general mechanism for tailoring light-matter interactions in low-loss dielectric metasurfaces and provide clear design principles for metasurface-based quantum emitters and nanophotonic devices operating in the telecom C-band.

\begin{figure}[ht]
    \centering
    \includegraphics[width=13cm]{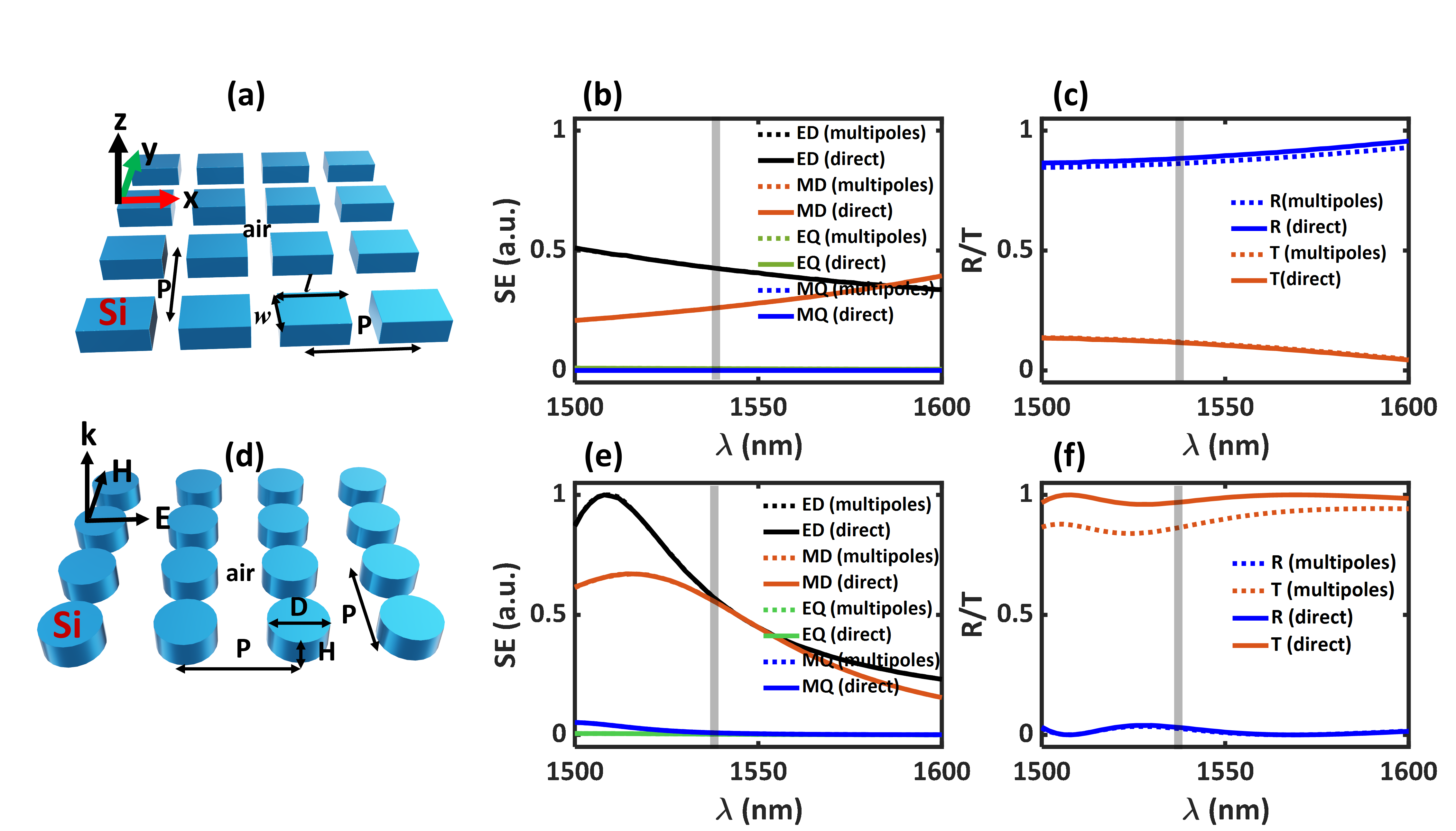}
    \caption{ (a) Schematic of a full silicon (Si) cuboid metasurface in air. (b) Spectral response of the multipolar modes under plane-wave excitation, showing good agreement between theoretical predictions (dotted lines) and numerical simulations (solid lines). (c) Reflectance and transmittance spectra of the cuboid metasurface obtained from theoretical analysis and direct full-wave simulations. (d) Schematic of the silicon disk metasurface in air. (e) Spectral response of the multipolar modes of the disk metasurface under plane-wave excitation, demonstrating close agreement between theoretical (dotted) and computational (solid) results. (f) Reflectance and transmittance spectra of the Si disk metasurface calculated using theoretical methods and validated by direct simulations.}
    \label{schematic_metasurface}
\end{figure}

\section{\label{sec:level2} \textit{Results and DISCUSSION} }

\subsection{Symmetric metasurfaces: multipolar response and validation}

We begin by analyzing the optical response of fully symmetric dielectric metasurfaces composed of silicon (Si) cuboid and disk meta-atoms, which serve as reference systems for assessing the impact of symmetry breaking. The geometries of the symmetric cuboid and disk metasurfaces are shown schematically in Fig.~\ref{schematic_metasurface}(a) and (d), respectively.\rev{ The geometrical parameters of the symmetric and symmetry-broken metasurfaces used in this work are summarized in Table~\ref{tab:geometry}.
\begin{table}[ht]
\centering
\caption{\revred{ Geometrical parameters of the symmetric and symmetry-broken metasurfaces.}}
\rev{\begin{tabular}{lcccccc}
\hline
Structure & $l$ (nm) & $w$ (nm) & $D$ (nm) & $h$ (nm) & $P$ (nm) & Asymmetry parameter \\
\hline
Symmetric cuboid metasurface & 526 & 410 & -- & 365 & 865 & -- \\
Broken cuboid metasurface & 526 & 410 & -- & 365 & 865 & $\delta l = 231$ nm, $\delta w = 204$ nm \\
Symmetric disk metasurface & -- & -- & 595 & 260 & 1000 & -- \\
Broken disk metasurface & -- & -- & 595 & 260 & 1000 & $\theta = 136^\circ$ \\
\hline
\end{tabular}}
\label{tab:geometry}
\end{table}

The geometric parameters were selected based on systematic parameter sweeps of the nanostructure (see \textit{Figs. S1 and S3 in supplementary section}).}

Under normal-incidence plane-wave excitation propagating along the $+z$ direction with the electric field polarized along the $+x$ direction, both metasurfaces support a hierarchy of Mie-type resonances. These include electric dipole (ED), magnetic dipole (MD), electric quadrupole (EQ), and magnetic quadrupole (MQ) modes excited within individual nanostructures \cite{allayarov2024, Alaee2018, Babicheva2017}. Fig.~\ref{schematic_metasurface}(b) and (e) show the wavelength-dependent scattering contributions of the dominant multipolar modes for the cuboid and disk metasurfaces, respectively. Excellent agreement is observed between the theoretical multipolar decomposition (dotted curves) and full-wave numerical simulations (solid curves), validating the accuracy of the multipolar framework. The corresponding reflectance and transmittance spectra, shown in Fig.~\ref{schematic_metasurface}(c) and (f), further confirm this agreement.
It can be seen here that neither symmetric metasurface exhibits a pronounced resonance in the vicinity of the target emission wavelength $\lambda = 1540$~nm. This absence of resonant enhancement at the emission wavelength highlights the necessity of symmetry breaking to enable efficient emitter-metasurface coupling.

\begin{figure}[ht]
    \centering
    \includegraphics[width=13cm]{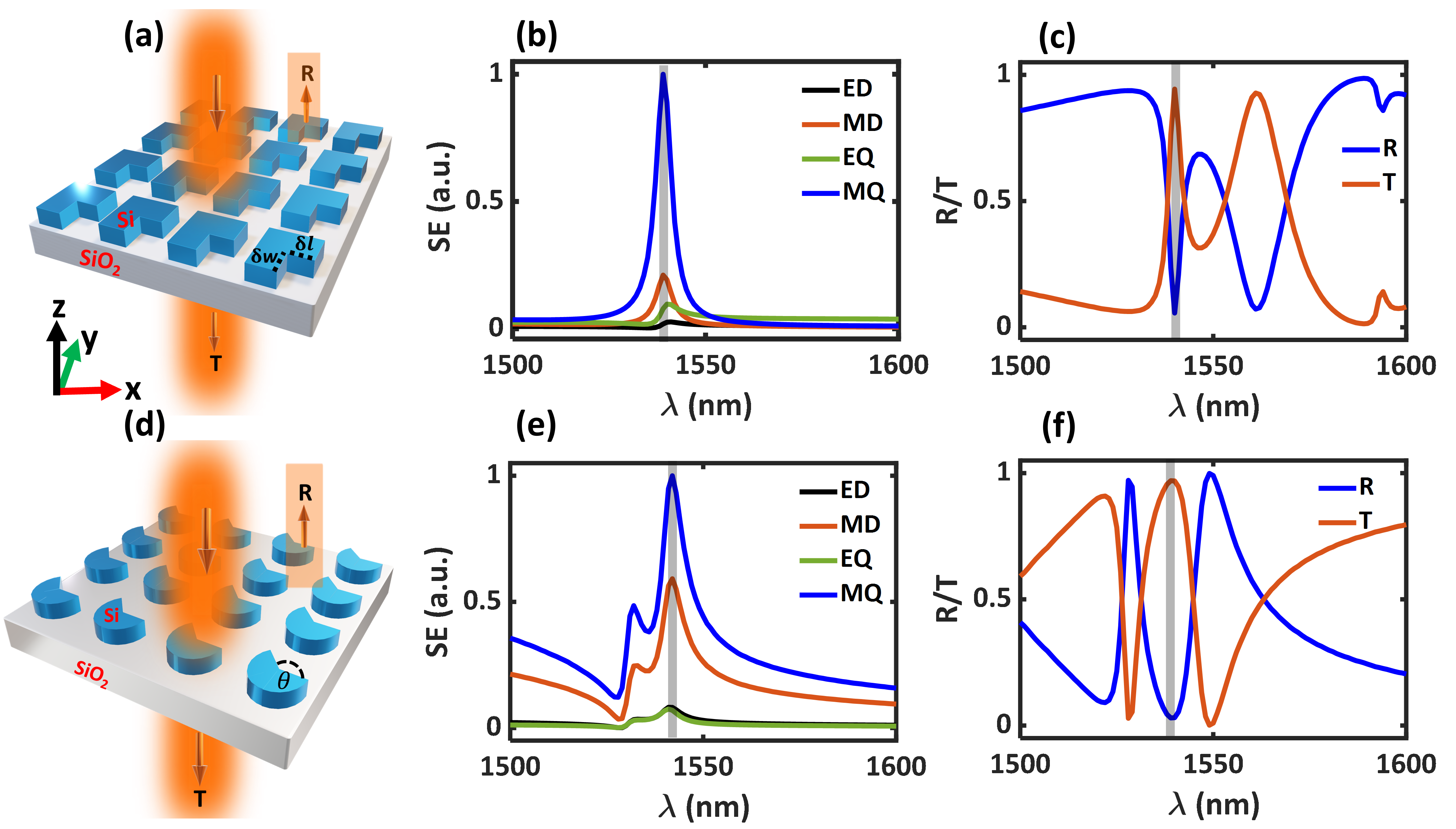}
    \caption{
(a,d) Schematic illustration of the reflectance ($R$) and transmittance ($T$) configurations of the broken Si cuboid and broken Si disk metasurfaces, respectively, under plane-wave excitation.
(b,e) Spectral response of the multipolar resonant modes of the broken cuboid and broken disk metasurfaces, respectively, under plane-wave excitation.
(c,f) Corresponding reflectance and transmittance spectra of the broken cuboid- and disk-based metasurfaces, respectively.
}

    \label{broken_modes}
\end{figure}

\begin{figure}[ht]
    \centering
\includegraphics[width=12cm]{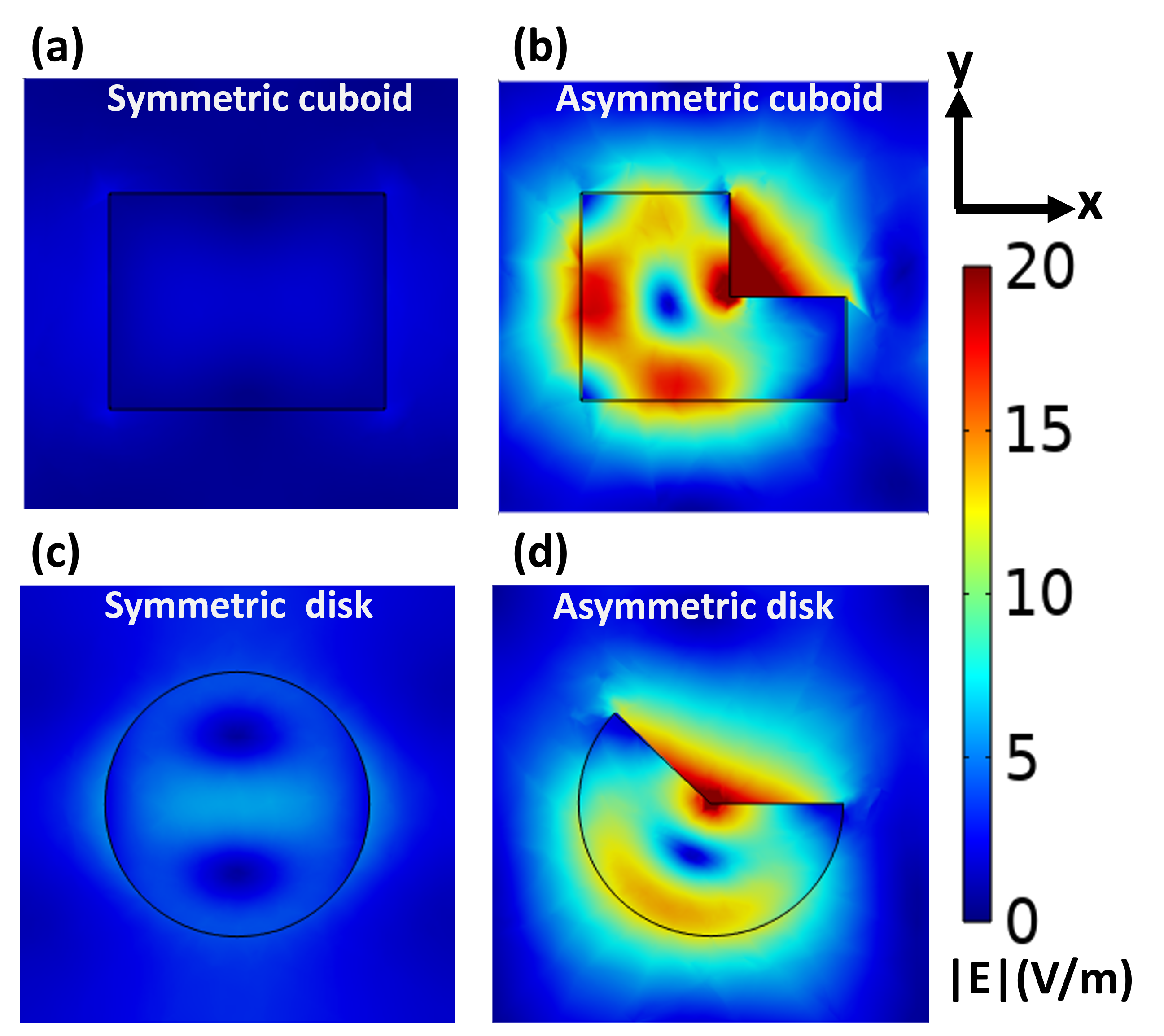}
    \caption{
Normalized 2D electric field profile distributions in the $xy$ plane taken (under plane wave excitation) at the vertical position of the embedded electric dipole for symmetric and symmetry-broken silicon cuboid (a and b) and (c and d) disk metasurfaces at $\lambda = 1540$~nm, respectively. Broken-symmetry geometries exhibit strong near-field localization due to magnetic multipolar hybridization, leading to enhanced electric-field confinement at the emitter position.}

    \label{field_profiles}
\end{figure}
\subsection{Symmetry-broken metasurfaces and multipolar hybridization}

We next investigate the effect of geometrical symmetry breaking on the metasurfaces' optical response. For the cuboid-based metasurface, asymmetry is introduced by removing a rectangular notch characterized by dimensions $\delta l$ and $\delta w$. In the disk-based metasurface, symmetry breaking is achieved by removing a sectorial segment defined by a cut angle $\theta$, as illustrated schematically in Fig.~\ref{broken_modes}(a) and (d) (see also \textit{ Supplementary Information Fig.~S1}).

By continuously varying the asymmetry parameters, symmetry-protected modes of the nanostructures are progressively hybridized, enabling coupling between otherwise orthogonal multipolar channels \cite{liu2018light,campione2016broken}. At optimized asymmetry values ($\delta l = 231~\mathrm{nm}$ and $\delta w = 204~\mathrm{nm}$  for the broken cuboid and at a cut angle $\theta = 136^{\circ}$ for the broken disk), both broken metasurfaces exhibit pronounced Fano-like spectral features arising from interference between bright and dark multipolar modes (\textit{ Supplementary Information Fig.~S1}).
\rev{These sharp spectral features are also consistent with the excitation of quasi-BIC-like resonances induced by symmetry breaking. Similar to previously reported symmetry-protected BIC metasurfaces, the geometrical perturbation opens a controlled radiative channel that enables coupling of otherwise weakly radiative modes to free space while preserving strong electromagnetic confinement \cite{li2019Bic,Shi2022}. In our case, however, the emphasis is not on chirality or polarization conversion, but on how this symmetry-enabled radiative access promotes MD--MQ hybridization and enhances the emitter-coupled near field.}

Fig.~\ref{broken_modes}(b) and (e) present the wavelength-dependent multipolar scattering contributions of the broken cuboid and disk metasurfaces, respectively. In both cases, the optical response at $\lambda = 1540$~nm is dominated by  MD and MQ modes. The corresponding reflectance and transmittance spectra, shown in Fig.~\ref{broken_modes}(c) and (f), display sharp resonant features at the same wavelength, confirming strong coupling between the incident field and the symmetry-broken metasurfaces.

These results demonstrate that geometrical asymmetry serves as a continuous tuning parameter that governs multipolar hybridization and resonance formation \cite {wu2023strong}.
The resonant multipolar hybridization induced by symmetry breaking directly affects the metasurfaces' near-field electromagnetic response. To elucidate this effect, Fig.~\ref{field_profiles} presents the normalized electric field profile in the $xy$ plane for symmetric and symmetry-broken cuboid and disk metasurfaces at the emission wavelength of $\lambda = 1540$~nm.

In the symmetric configurations, the electric field remains weakly localized within the nanostructures [Figs.~\ref{field_profiles}(a) and \ref{field_profiles}(c)], consistent with the absence of strong resonant excitation of the dominant multipolar modes at this wavelength [Figs.~\ref{schematic_metasurface}]

In contrast, the symmetry-broken metasurfaces exhibit pronounced near-field localization [Figs.~\ref{field_profiles}(b) and \ref{field_profiles}(d)]. This enhanced field confinement originates from the hybridization of MD and MQ modes enabled by symmetry breaking, which shifts their resonant response toward $\lambda = 1540$~nm. The resulting magnetic multipolar interference leads to a strongly enhanced electric field near the center of the unit cell, providing favorable conditions for efficient emitter-metasurface coupling.

\begin{figure}[ht]
    \centering
   \includegraphics[width=13cm]{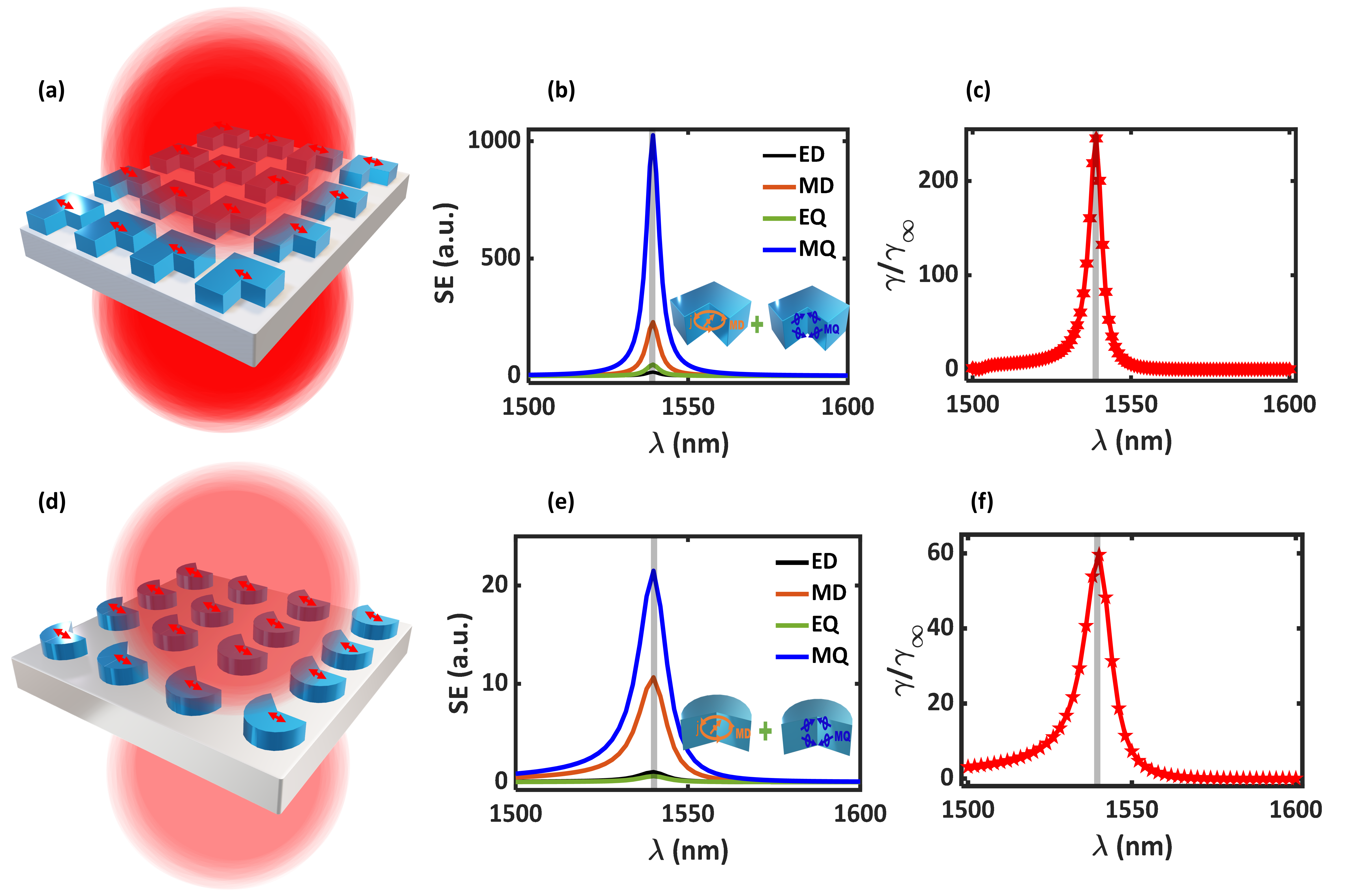}
    \caption{
(a) and (d) Schematic illustration of enhanced dipole emission from a broken cuboid and broken disk metasurface, where each unit cell is embedded with a horizontally aligned electric dipole near its center.
(b) and (e) Spectral response of the dominant electromagnetic resonant modes in the broken cuboid and disk-based metasurfaces, respectively, showing enhancement in the SE compared to their symmetric counterpart.
(c) and (f) Normalized decay rate $\gamma/\gamma_{\infty}$ of a dipole embedded at the center of each broken cuboid and disk-based metasurface as a function of wavelength $\lambda$, respectively, exhibiting enhancements of up to two orders of magnitude for broken cuboid and an order of magnitude for broken disk at $\lambda: 1540$ nm.}

    \label{decay_rt}
\end{figure}

\subsection{ Spontaneous emission enhancement}
To investigate the impact of symmetry breaking on spontaneous emission, we embed a classical electric dipole emitter near the center of each unit cell of the broken cuboid and disk metasurfaces. The dipole is oriented horizontally and positioned at the region of maximum modal field confinement to ensure optimal coupling to the resonant eigenmodes of the metasurface, as shown in Figs.~\ref{decay_rt}(a) and \ref{decay_rt}(d).

Within the framework of macroscopic quantum electrodynamics, the spontaneous emission rate of a dipole at position $\mathbf{r}$ and frequency $\omega$ is governed by the photonic local density of states (LDOS), which is directly proportional to the imaginary part of the electromagnetic Green’s tensor evaluated at the emitter position \cite{novotny2012principles}:
\begin{equation}
\rho(\mathbf{r},\omega)=\frac{2\omega}{\pi c^2}
\mathrm{Im}\left[\mathbf{p}^*\cdot
\mathbf{G}(\mathbf{r},\mathbf{r},\omega)\cdot
\mathbf{p}\right],
\end{equation}
where $\mathbf{G}(\mathbf{r},\mathbf{r},\omega)$ is the dyadic Green’s function of the structured photonic environment and $\mathbf{p}$ denotes the dipole moment orientation. The normalized decay rate is therefore given by
\begin{equation}
\frac{\gamma}{\gamma_\infty}=
\frac{
\mathrm{Im}\left[\mathbf{p}^*\cdot
\mathbf{G}(\mathbf{r},\mathbf{r},\omega)\cdot
\mathbf{p}\right]
}{
\mathrm{Im}\left[\mathbf{p}^*\cdot
\mathbf{G}_\infty(\mathbf{r},\mathbf{r},\omega)\cdot
\mathbf{p}\right]
},
\end{equation}
where $\mathbf{G}_\infty$ corresponds to Green’s tensor in the reference system.

%The total Green’s tensor of the metasurface can be decomposed as 
%$\mathbf{G}=\mathbf{G}_0+\mathbf{G}_{\mathrm{scat}}$, 
%where $\mathbf{G}_{\mathrm{scat}}$ represents the scattering contribution arising from the resonant response of the structure. Hence, any modification of the spontaneous emission rate originates from changes in the scattered electromagnetic field fed back to the emitter.

In the broken metasurfaces, the dipole excitation induces hybridized magnetic dipole (MD) and magnetic quadrupole (MQ) resonances at $\lambda=1540$~nm [Figs.~\ref{decay_rt}(b) and \ref{decay_rt}(e)]. Symmetry breaking enables coherent coupling between these otherwise orthogonal multipolar channels, leading to strong modal interference and enhanced resonant excitation. As a direct consequence, the scattering efficiency of the induced MD and MQ modes increases by nearly three orders of magnitude for the cuboid geometry and by more than one order of magnitude for the broken disk relative to their symmetric counterparts.

The strengthened multipolar scattering amplifies the scattered near field at the emitter position, thereby enhancing the imaginary part of the  Green’s tensor. This results in a substantial increase of the LDOS and, consequently, the normalized spontaneous emission rate. \revred{At the target wavelength of 1540~nm, the symmetric cuboid and symmetric disk metasurfaces exhibit normalized decay rates of only $1.441$ and $6.602$, respectively. By comparison, the broken-cuboid and broken-disk metasurfaces yield $\gamma/\gamma_\infty \approx 268.5$ and $\approx 60$, corresponding to enhancement factors of approximately $186.3$ and $9.1$, respectively, relative to their symmetric counterparts. These comparisons show explicitly that the strong enhancement arises from symmetry-breaking-induced multipolar hybridization rather than from the silicon metasurface environment alone.}

Thus, symmetry-enabled coherent multipolar interaction modifies the electromagnetic response function of the environment through enhanced resonant scattering, providing deterministic control over the photonic LDOS and spontaneous emission dynamics in low-loss dielectric metasurfaces.

\subsection{\revred{Finite-array validation and radiative efficiency}}

\begin{figure}[ht]
    \centering
    \includegraphics[width=15cm]{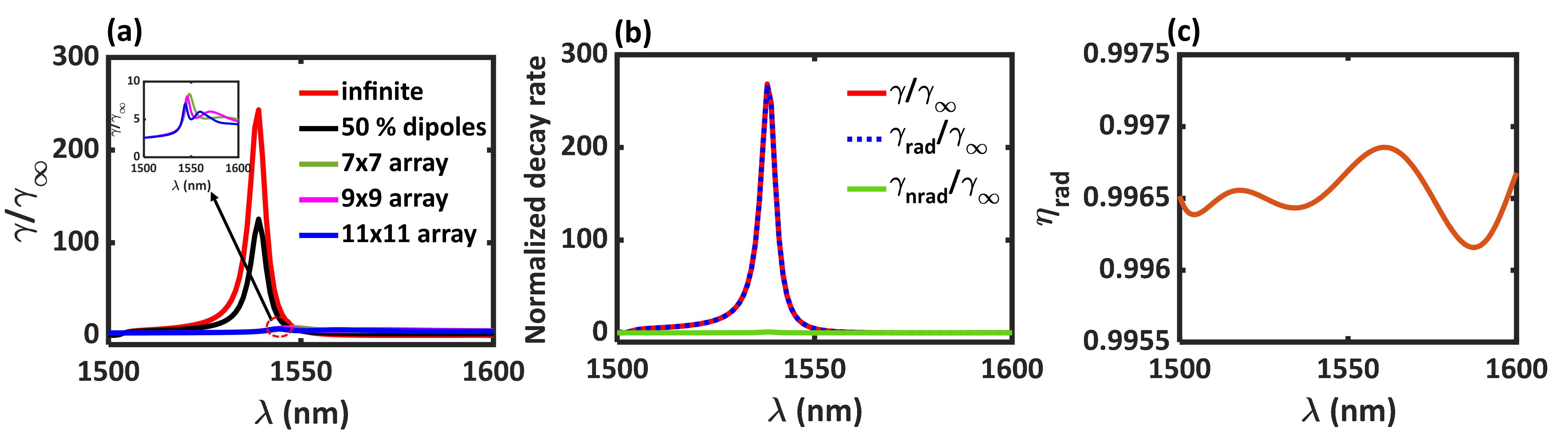}
    \caption{\revred{Validation of the decay-rate enhancement against artificial source repetition. The main panel compares the normalized decay-rate spectra of the fully periodic broken-cuboid metasurface (infinite lattice with one dipole per unit cell), a periodic $2\times2$ supercell with 50\% dipole occupancy, and finite $7\times7$, $9\times9$, and $11\times11$ broken-cuboid arrays containing only a single dipole in the central resonator. The finite-array calculations yield substantially smaller enhancement than the fully periodic case and exhibit a progressive shift of the resonance wavelength toward 1540~nm as the array size increases. The inset highlights the finite-array spectra near resonance. (b) Spectral variation of the total, radiative, and non-radiative decay-rate enhancements, $\gamma/\gamma_\infty$, $\gamma_{\mathrm{rad}}/\gamma_\infty$, and $\gamma_{\mathrm{nrad}}/\gamma_\infty$, respectively. The close overlap of $\gamma$ and $\gamma_{\mathrm{rad}}$ shows that the enhancement is predominantly radiative, while the non-radiative contribution remains small. (c) Corresponding radiative efficiency $\eta_{\mathrm{rad}}=\gamma_{\mathrm{rad}}/\gamma$ as a function of wavelength. The efficiency remains close to unity throughout the resonance region, reaching approximately $0.9966$ at 1540~nm.}}
    \label{finite_array_validation_fig}
\end{figure}

\revred{Because a unit-cell calculation with periodic boundary conditions corresponds to an idealized two-dimensional phased array of perfectly coherent and identical emitters, we further examined how the decay-rate enhancement changes when this collective source repetition is reduced. Figure~\ref{finite_array_validation_fig}(a) compares the normalized decay-rate spectra for the fully periodic broken-cuboid metasurface, a periodic $2\times2$ supercell with 50\% dipole occupancy, and finite $7\times7$, $9\times9$, and $11\times11$ broken-cuboid arrays containing only a single dipole in the central resonator. The fully periodic case yields $\gamma/\gamma_\infty \approx 268$-fold near 1540~nm, whereas the 50\% dipole supercell gives a reduced enhancement of about $125$ fold, already indicating the importance of coherent source repetition. For the finite arrays, the peak enhancement decreases further to about $9$-fold at 1548~nm for the $7\times7$ array, $8$-fold at 1546~nm for the $9\times9$ array, and $7$-fold at 1544~nm for the $11\times11$ array. At the same time, the resonance wavelength shifts progressively toward the periodic-lattice value of 1540~nm as the array size increases. This trend indicates convergence from an edge-influenced finite-array response toward the intrinsic isolated-emitter lattice response and confirms that the very large PBC value contains a substantial collective contribution associated with coherent emitter repetition. Owing to memory limitations, the finite-array calculations were carried out up to the $11\times11$ case, which already shows a clear convergence trend.}

\revred{To distinguish radiative decay from non-radiative quenching, we further separated the total, radiative, and non-radiative decay-rate contributions as shown in Fig~\ref{finite_array_validation_fig}(b). At 1540~nm, the calculated values are $\gamma/\gamma_\infty = 268.4844$, $\gamma_{\mathrm{rad}}/\gamma_\infty = 267.5672$, and $\gamma_{\mathrm{nrad}}/\gamma_\infty = 0.9172$, so that the non-radiative contribution remains very small compared with the total enhanced decay. The corresponding radiative efficiency is $\eta_{\mathrm{rad}}=\gamma_{\mathrm{rad}}/\gamma\approx 0.9966$ (Fig~\ref{finite_array_validation_fig}(c)), showing that within the present electrodynamic model the enhancement is overwhelmingly radiative and is not dominated by external non-radiative quenching in the photonic structure.}

\begin{figure}[ht]
    \centering
    \includegraphics[width=13cm]{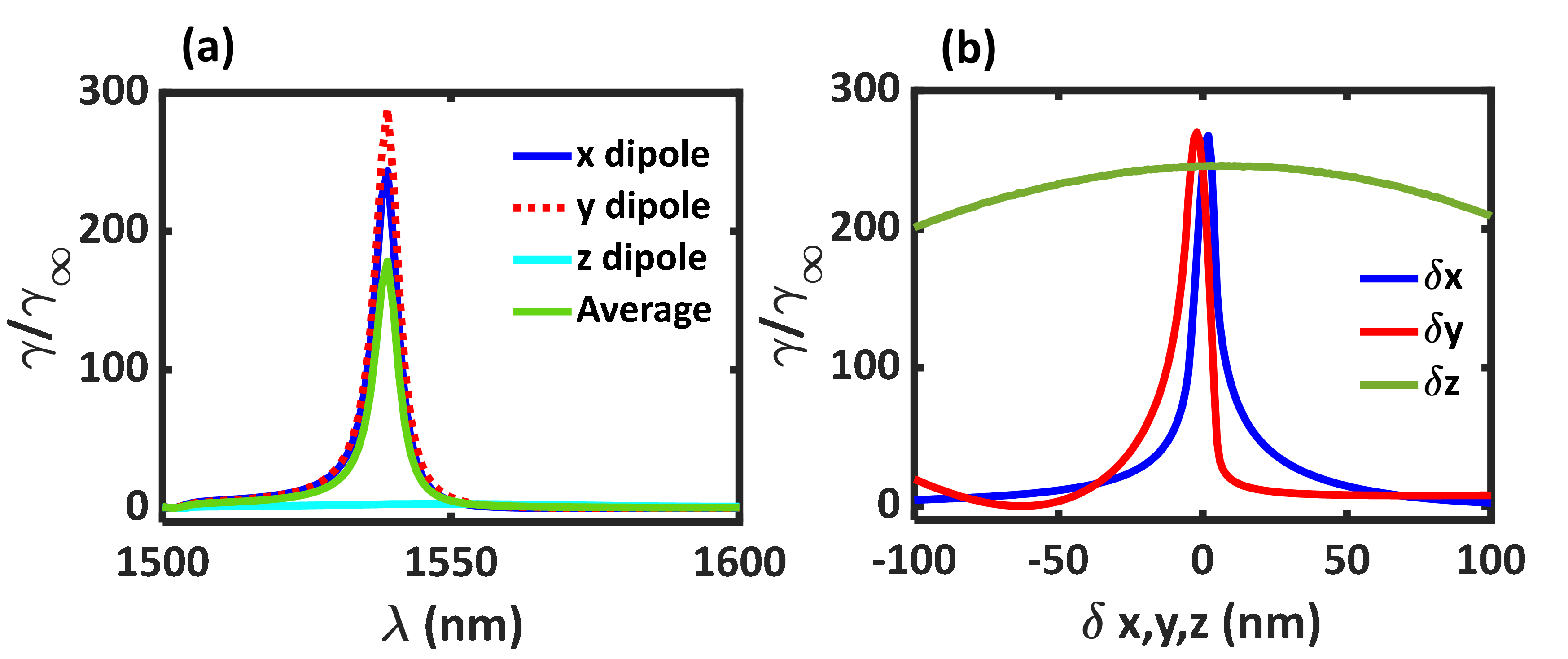}
    \caption{\revred{Orientation and spatial-tolerance analysis of the spontaneous-emission enhancement in the broken-cuboid metasurface. (a) Normalized decay-rate enhancement $\gamma/\gamma_{\infty}$ as a function of wavelength for $x$-, $y$-, and $z$-oriented dipoles located at the optimized hotspot position. Strong enhancement is obtained for the in-plane dipole orientations, whereas the $z$-oriented dipole couples much more weakly to the resonant mode. The bold green curve represents the simple orientational average $(\gamma_x+\gamma_y+\gamma_z)/3$. (b) Normalized decay-rate enhancement at the resonance wavelength as a function of emitter displacement $\delta x$, $\delta y$, and $\delta z$ from the optimized hotspot position. The enhancement is strongly localized for lateral displacements and varies more gradually along the vertical direction over the scanned range.}}
    \label{orientation_tolerance_fig}
\end{figure}

\subsection{\revred{Dipole orientation, spatial tolerance, and spectral bandwidth}}
\label{subsec:dipole_orientation}

\revred{Because realistic Er$^{3+}$ emitters are not perfectly aligned or positioned exactly at the electromagnetic hotspot, we further examined the influence of dipole orientation and emitter displacement on the calculated enhancement for the broken-cuboid metasurface. Fig.~\ref{orientation_tolerance_fig}(a) compares the normalized decay-rate spectra for $x$-, $y$-, and $z$-oriented dipoles placed at the optimized hotspot position. Strong enhancement is obtained for the two in-plane orientations ($x$ and $y$), whereas the $z$-oriented dipole couples much more weakly to the resonant mode. This behavior reflects the vector character of the resonant near field: at the hotspot, the MD--MQ hybridized mode produces predominantly in-plane electric-field components, so dipoles polarized along $x$ and $y$ couple efficiently to the mode, while a $z$-oriented dipole experiences much weaker overlap. Therefore, the maximum decay-rate enhancement reported in this work should be interpreted as the optimal value for an in-plane ($x$- and $y$-oriented) emitter. For a randomly oriented ensemble, the effective enhancement would be reduced relative to the optimal in-plane case; a simple orientational average may be estimated as $(\gamma_x+\gamma_y+\gamma_z)/3$, where $\gamma_x$, $\gamma_y$, and $\gamma_z$ are the normalized decay-rate enhancements for $x$-, $y$-, and $z$-oriented dipoles, respectively highlighted by the bold green curve in Fig.~\ref{orientation_tolerance_fig} (a).}

\revred{Fig.~\ref{orientation_tolerance_fig}(b) shows the dependence of the enhancement on emitter displacement from the hotspot along the $x$, $y$, and $z$ directions, where the emitter position was varied over the range of $\pm 100$~nm about the optimized hotspot. The enhancement decreases more rapidly for lateral displacements $\delta x$ and $\delta y$, confirming that the resonant field is strongly localized in the metasurface plane. By contrast, the variation with $\delta z$ is more gradual over the scanned range, indicating weaker sensitivity along the vertical direction. Physically, this difference arises because the MD--MQ hybridized mode forms a tightly confined in-plane hotspot near the center of the broken resonator, so lateral displacements reduce the emitter--mode overlap more strongly than vertical displacement. Along the vertical direction, however, the modal field extends more gradually away from the resonator plane, leading to a slower reduction of the LDOS enhancement. It is also noted that the $\delta y$ dependence is not perfectly symmetric about the nominal hotspot position and that its maximum is slightly shifted relative to that of $\delta x$. This behavior originates from the broken structural symmetry of the cuboid, which produces an anisotropic near-field distribution and shifts the local field maximum away from the geometric center. As a result, the field gradient along the $y$ direction is different on the two sides of the hotspot, leading to a sharper variation on one side and a peak that need not coincide exactly with $\delta x=0$. These results show that the peak enhancement should be regarded as an upper-bound value for an optimally positioned emitter, whereas experimentally relevant enhancement remains finite over a nonzero spatial region around the hotspot.}

 \revred{\revred{Finally, the large peak enhancement is associated with a spectrally sharp MD-MQ hybridized resonance centered near the target Er$^{3+}$ emission wavelength of $1540$~nm. From the calculated spectra of the broken-cuboid metasurface, the dominant MD and MQ resonances exhibit full widths at half maximum of approximately $5.0$~nm, while the corresponding decay-rate enhancement peak has a full width at half maximum of approximately $4.7$~nm. Room-temperature Er$^{3+}$ emission near $1.54~\mu$m is known to exhibit a finite linewidth of the order of $10$~nm in silicon-based devices \cite{zheng1994Er}. Therefore, the reported peak decay-rate enhancement should be interpreted as the ideally spectrally matched limit. For a realistic room-temperature Er$^{3+}$ ensemble, only the spectral portion overlapping with the MD-MQ resonance would experience the full peak enhancement, and the effective enhancement after spectral averaging over the emitter linewidth would be reduced. Nevertheless, because the resonance remains centered close to the target Er$^{3+}$ emission wavelength, significant spectral overlap is retained. A similar consideration applies to the broken-disk metasurface, for which the peak value should likewise be interpreted as the optimal spectrally matched enhancement rather than a full ensemble-averaged value.}}

% \revred{Finally, the large peak enhancement is associated with a spectrally sharp MD--MQ hybridized resonance. Room-temperature Er$^{3+}$ emission near 1.54~$\mu$m is known to exhibit a finite linewidth of the order of $10$~nm in silicon-based devices \cite{zheng1994Er}. Therefore, the reported peak decay-rate enhancement corresponds to the ideally spectrally matched case, while the effective enhancement for a realistic room-temperature Er$^{3+}$ ensemble would be reduced after spectral averaging over the finite emitter linewidth. A similar consideration applies to the broken-disk metasurface, for which the peak value should likewise be interpreted as the optimal spectrally matched enhancement rather than a full ensemble-averaged value.}

\subsection{\revred{Fabrication tolerance and practical feasibility}}

\begin{figure}[ht]
\centering
\includegraphics[width=15cm]{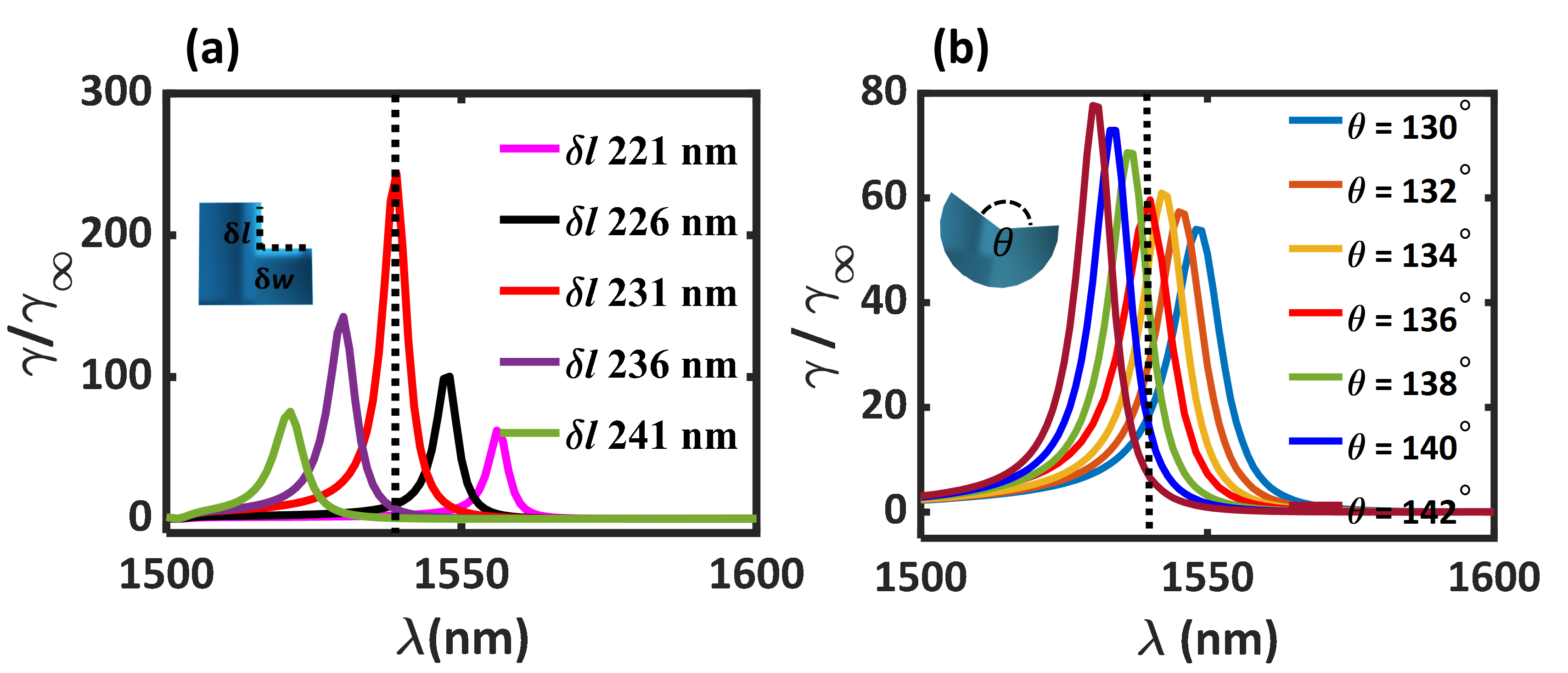}
\caption{\revred{Fabrication-tolerance analysis of the optimized symmetry-broken metasurfaces. (a) Normalized decay-rate enhancement $\gamma/\gamma_{\infty}$ of the broken-cuboid metasurface for different notch lengths $\delta l$ around the optimized value of $231$~nm. (b) Normalized decay-rate enhancement $\gamma/\gamma_{\infty}$ of the broken-disk metasurface for different sector cut angles $\theta$ around the optimized value of $136^\circ$. In both cases, moderate geometrical deviations preserve the resonant enhancement but lead to spectral detuning and variation in the peak enhancement, indicating that the underlying MD--MQ hybridization mechanism is robust while exact spectral alignment still requires controlled fabrication.}}
\label{tolerance_fig}
\end{figure}

\revred{To assess the practical robustness of the optimized symmetry-broken metasurfaces, we performed additional tolerance analyses by varying the geometrical asymmetry parameters around their optimized values and monitoring the resulting decay-rate spectra. For the broken-cuboid metasurface, the notch length was varied over the range $\delta l = 221$--$241$~nm around the optimized value of $231$~nm. For the broken-disk metasurface, the sector cut angle was varied over the range $\theta = 130^\circ$--$142^\circ$ around the optimized value of $136^\circ$. The corresponding spectra are summarized in Fig.~\ref{tolerance_fig}, while the effect of notch-width variation $\delta w$ for the broken-cuboid metasurface is provided in the Supplementary Information.}

\revred{Fig.~\ref{tolerance_fig}(a) shows that the broken-cuboid metasurface retains a clear resonant enhancement over the investigated range of notch lengths $\delta l$, although the spectral position and magnitude of the peak vary as the geometry is detuned from the optimized value. The strongest enhancement is obtained near $\delta l = 231$~nm, while deviations from this value shift the resonance and reduce the peak decay-rate enhancement. The influence of notch-width variation $\delta w$ exhibits a qualitatively similar behavior and is discussed in the Supplementary Information (\textit{Fig. S4}).}

\revred{Fig.~\ref{tolerance_fig}(b) shows that the broken-disk metasurface also preserves its resonant enhancement over the investigated range of cut angles $\theta$. In this case, the resonance wavelength shifts systematically as $\theta$ is varied, while the peak enhancement changes in a non-monotonic manner about the optimized value. The highest enhancement and best spectral alignment with the target Er$^{3+}$ emission wavelength occur near the optimized geometry, whereas deviations from this value primarily detune the resonance rather than eliminating it.}

\revred{These results indicate that the underlying symmetry-breaking-induced MD--MQ hybridization mechanism is moderately robust against fabrication imperfections. In both metasurfaces, the dominant consequence of geometrical fluctuation is spectral detuning rather than a disappearance of the enhancement mechanism. Therefore, while precise control of the asymmetry parameters is required to align the resonance accurately with the Er$^{3+}$ emission wavelength at 1540~nm, the proposed metasurface designs remain practically meaningful under moderate fabrication deviations. }
\revred{From the standpoint of practical implementation, the relevant structural dimension (of the order of 200 nm) remains within the range accessible to standard silicon nanofabrication technologies.
Such dimensions are, in principle, compatible with established top-down fabrication routes such as electron-beam lithography or other high-resolution lithographic patterning followed by reactive-ion etching. Importantly, the broken-symmetry features considered here do not rely on extremely small sub-10-nm details, which supports the practical feasibility of the proposed geometries.}

\begin{figure}[ht]
    \centering
   \includegraphics[width=13cm]{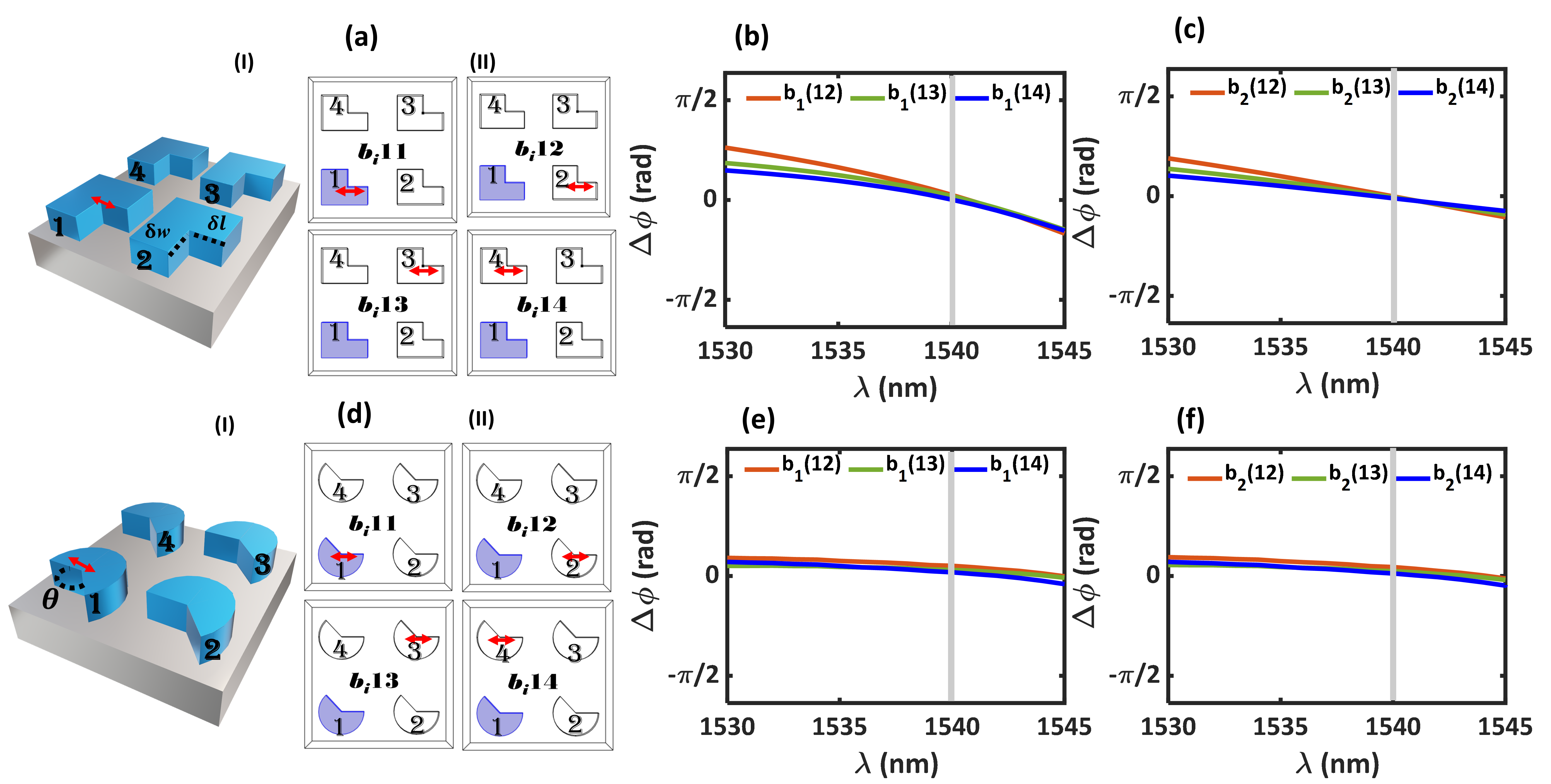}
    \caption{
(a) and (d) ( (I) Schematic illustration of the four-element broken cuboid and disk-based unit cell.
(II) Schematic of the phase analysis of the magnetic dipole (MD) $b_{1}(1i)$ and magnetic quadrupole (MQ) $b_{2}(1j)$ modes, where $i,j=1,2,3,4$, arising from interactions among the surrounding cuboids and disk when an embedded dipole is present in them, respectively.)
(b) and (e) Phase difference $\Delta\phi$ as a function of  $\lambda$ for MD mode interactions among the surrounding broken cuboids and disk, respectively.
(c) and (f) Phase difference $\Delta\phi$ as a function of $\lambda$ for MQ mode interactions among the surrounding broken cuboids and disk, respectively.}
%\revred{The phases are extracted from the arguments of the complex volume-integrated MD and MQ multipole amplitudes of neighboring resonators within the periodic four-element supercell.}}
  
    \label{Phases}
\end{figure}

\subsection{Phase-resolved multipolar coherent interaction}

In order to explain the physical mechanism governing the observed enhancement, we perform a phase-resolved analysis of the dominant multipolar responses using a metasurface supercell composed of four neighboring elements, as illustrated in Figs.~\ref{Phases}(a) and \ref{Phases}(d). An electric dipole emitter is sequentially embedded within each element, and the phase evolution of the induced magnetic dipole ($b_1$) and magnetic quadrupole ($b_2$)  moments is evaluated to probe inter-element coupling mediated by near-field interactions.

\revred{To address the methodological scope of this analysis, we note that the four-element configuration was chosen as the smallest periodic supercell that allows explicit comparison of inter-element phase relations between neighboring resonators. A single unit cell does not permit such relative phase extraction, whereas the four-element periodic supercell provides a compact and computationally tractable geometry for resolving nearest-neighbor phase coherence while preserving the periodic character of the metasurface. Periodic boundary conditions (PBCs) are applied on the lateral boundaries of this supercell, while the out-of-plane boundaries are treated consistently with the main periodic metasurface model. The phase is extracted from the complex induced current moments, not from scattered-field coefficients. Specifically, the MD and MQ phases are obtained from the arguments of the corresponding complex volume-integrated multipole amplitudes calculated from the induced polarization/current distribution inside each resonator. The plotted phase differences are defined as the relative phase differences between neighboring elements for the same multipolar channel.}

Figs.~\ref{Phases}(b) and \ref{Phases}(e) illustrate the wavelength-dependent relative phase evolution of MD modes in neighboring unit cells of the broken cuboid and disk metasurfaces, respectively, while Figs.~\ref{Phases}(c) and \ref{Phases}(f), the corresponding phase behavior of the magnetic quadrupole (MQ) modes (see Supplementary Fig.~S2 for asymmetry dependence phase analysis).
 In both geometries, the induced MD modes exhibit near-zero phase difference across the supercell at $\lambda = 1540$ nm, indicating phase-coherent excitation. A similar phase coherence is observed for the MQ modes at the same wavelength.

% \rev{The simultaneous phase coherence of both MD and MQ modes at $\lambda = 1540$~nm indicates that these multipolar modes are excited in a mutually coherent manner across the metasurface. The phase analysis employed here provides a multipolar interpretation of the near-field interaction responsible for LDOS modification and spontaneous-emission enhancement in symmetry-broken metasurfaces. This picture is also compatible with the broader interpretation of symmetry-broken high-$Q$ resonances in dielectric metasurfaces, where opening a radiative channel preserves strong modal confinement while enabling efficient coupling to external fields \cite{li2019Bic,Shi2022,Hong2025single}. In the present structures, the physically relevant consequence is the phase-coherent MD--MQ interaction that governs the emitter-coupled near field, rather than a chiroptical response.}
\rev{The simultaneous phase coherence of both MD and MQ modes at $\lambda = 1540$~nm indicates that these multipolar modes are excited in a mutually coherent manner across the metasurface. The phase analysis employed here is used as an interpretive tool to describe the near-field multipolar interaction responsible for LDOS modification and spontaneous-emission enhancement in the present symmetry-broken metasurfaces. This interpretation is consistent with the broader picture of symmetry-broken high-$Q$ resonances in dielectric metasurfaces, where a geometrical perturbation opens a radiative channel for otherwise weakly radiative modes while preserving strong modal confinement \cite{koshelev2018assymetric,li2019Bic}. Related symmetry-broken and chiral dielectric metasurfaces have also used quasi-BIC physics to realize polarization-selective or chiroptical resonances \cite{Shi2022,Hong2025single}. In the present structures, however, the physically relevant consequence is the phase-coherent MD--MQ interaction that governs the emitter-coupled near field, rather than a chiroptical response.} When combined with their spectral overlap and enhanced amplitudes, this coherence enables constructive near-field interference at the emitter position. Thus, this coherent multipolar interaction enabled by symmetry-breaking in the metasurfaces results in strong near-field localization at the emitter position, which enhances the LDOS and leads to an increase in the spontaneous emission.

\section{\label{sec:level3} \textit{METHODS} }
\subsection{Metasurfaces modeling}\label{sec9}
All electrodynamical simulations were performed using the finite-element method (FEM) implemented in the RF module of COMSOL Multiphysics. The quantum emitter is modeled as a radiating point electric dipole, represented by an oscillating point current source operating at the emission wavelength $\lambda$ \cite{xu1999finite,novotny2012principles}. A periodic array of symmetry-broken metasurfaces is realized by applying periodic boundary conditions (PBCs) on the four lateral faces of the simulation domain. The top and bottom boundaries are terminated using second-order scattering/perfectly matched layers (PMLs) boundary conditions to suppress artificial reflections. The computational mesh is refined with a minimum element size of $1~\mathrm{nm}$ and a maximum element size of $\lambda/7$ to ensure numerical accuracy. The wavelength-dependent permittivity values of Si and $\mathrm{SiO_2}$ are taken from experimentally reported data \cite{10.1063/1.555624,Malitson:65}.

\subsection{Multipolar decomposition and optical response}

Dielectric nanostructures with characteristic dimensions comparable to the wavelength of electromagnetic radiation act as efficient optical scatterers. When illuminated by an external electromagnetic field, such structures support induced charge and current distributions within their volume. In the frequency domain, the induced current density $\mathbf{J}_{\omega}(\mathbf{r})$ is expressed as \cite{hinamoto2021menp}
\begin{equation}
\mathbf{J}_{\omega}(\mathbf{r}) = i\omega \epsilon_0 \left(\epsilon_r - 1\right)\mathbf{E}_{\omega}(\mathbf{r}),
\end{equation}
where $\epsilon_0$ and $\epsilon_r$ are the permittivity of free space and the relative permittivity of the scattering medium, respectively, and $\mathbf{E}_{\omega}(\mathbf{r})$ denotes the electric field inside the dielectric nanostructure.

The induced current density acts as the fundamental source of Mie-type resonances, giving rise to a hierarchy of multipolar scattering modes, including electric dipole (ED), magnetic dipole (MD), electric quadrupole (EQ), magnetic quadrupole (MQ), and higher-order moments \cite{Bohren1998, alaee2018electromagnetic}. The total scattering  response of a metasurface unit cell is governed by the relative strength and interference of these multipolar contributions and can be quantified through the total scattering efficiency (SE), given by \cite{alaee2018electromagnetic}
\begin{subequations}\label{Scattering_Eq}
\begin{align}
C_{sca}^{total} &= C_{sca}^{p}+C_{sca}^{m}+C_{sca}^{Q}+C_{sca}^{M}, \\
C_{sca}^{total} &= \frac{k^4}{6\pi \epsilon_0^2 |E_{inc}|^2}
\Biggl\{
\sum \left( |p_{\alpha}|^2 + \left| \frac{m_{\alpha}}{c} \right|^2 \right)
+ \frac{1}{120}
\sum \left( |kQ_{\alpha\beta}^{e}|^2 + \left| \frac{kQ_{\alpha\beta}^{m}}{c} \right|^2 \right)
\Biggr\},
\end{align}
\end{subequations}
where $E_{inc}$ denotes the amplitude of the incident electric field, $k$ is the free-space wave vector, and $c$ is the speed of light.

Here, $p_{\alpha}$ and $m_{\alpha}$ represent the Cartesian components of the electric and magnetic dipole moments, while $Q_{\alpha\beta}^{e}$ and $Q_{\alpha\beta}^{m}$ correspond to the electric and magnetic quadrupole tensors, respectively. These multipolar moments are obtained from the induced current density according to \cite{alaee2018electromagnetic}
\begin{subequations}
\begin{align}
ED:p_{\alpha} & = -\frac{1}{i\omega}\Biggl\{\int{d}^{3}\mathbf{r}{J}_{\alpha}^{\omega}j_{0}(kr)+\frac{k^2}{2}\int{d}^{3}\mathbf{r}\Big[ 3(\mathbf{r}.\mathbf{J}_{\omega}){r}_{\alpha} \nonumber\\ & -{r}^2{J}_{\alpha}^{\omega} \Bigr] \frac{j_{2}(kr)}{(kr)^2}\Biggr\} \\
MD:m_{\alpha} & = \frac{3}{2}\int{d}^{3}\mathbf{r}(\mathbf{r}\times \mathbf{J}_{\omega})_{\alpha}\frac{j_{1}(kr)}{kr} \\
EQ:Q_{\alpha\beta}^{e} & = -\frac{3}{i\omega}\Biggl\{\int{d}^{3}\mathbf{r}[3(r_{\beta}{J}_{\alpha}^{\omega}+r_{\alpha}{J}_{\beta}^{\omega}) \nonumber\\ & -2(\mathbf{r}.\mathbf{J}_{\omega}){\delta}_{\alpha\beta}]\frac{j_{1}(kr)}{(kr)} +  2k^2\int{d}^{3}r[5r_{\alpha}r_{\beta}(\mathbf{r}.\mathbf{J}_{\omega}) \nonumber\\ & -(r_{\alpha}J_{\beta}+r_{\beta}J_{\alpha})r^2 -  r^2(\mathbf{r}.\mathbf{J}_{\omega}){\delta}_{\alpha\beta}]\frac{j_{3}(kr)}{(kr)^3}\Biggr\}  \\
MQ:Q_{\alpha\beta}^{m} & = 15\int{d}^{3}\mathbf{r}\Bigl\{r_{\alpha}(\mathbf{r}\times \mathbf{J}_{\omega})_{\beta} \nonumber\\ & + r_{\beta}(\mathbf{r}\times \mathbf{J}_{\omega})_{\alpha}\Bigr\}\frac{j_{2}(kr)}{{(kr)^2 }}
\end{align}
\label{Partial_SE}
\end{subequations}

All volume integrals in Eq.~\eqref{Partial_SE} are evaluated over the scattering medium, corresponding to the silicon cuboid and disk constituting the metasurface unit cell in the present work.

\revred{For the phase-resolved analysis discussed in Section~\ref{sec:level2}, the electric and magnetic multipole components were constructed in COMSOL from the induced polarization/current variables using the exact multipole-expansion expressions of Eq.~\eqref{Partial_SE}. These quantities are then integrated over the resonator volume to obtain the corresponding complex multipole amplitudes. The phase of a given multipolar mode is taken as the argument of the corresponding complex integrated amplitude\cite{Shamkhi2019}. In practice, the dominant MD and MQ phases were obtained from the arguments of the corresponding complex multipolar amplitudes, and the relative phase difference was evaluated between neighboring resonators for the same multipolar mode. Accordingly, the phase analysis reported in Fig.~\ref{Phases} is based on complex integrated multipole amplitudes derived from induced current/polarization moments.}

The reflectance and transmittance coefficients of the metasurfaces are subsequently evaluated using the contributions of the excited multipolar resonances, following the formalism described in Refs.~\cite{alaee2018electromagnetic,terekhov2019multipole}. Within this framework, the complex transmission coefficient $t$ can be expressed as
\begin{subequations}\label{RandT}
\begin{align}
t = 1 + \frac{ik_{d}}{2E_{0}S_{L}\epsilon_{0}\epsilon_{d}}
\left(
p_{x} + \frac{m_{y}}{v_{d}}
- \frac{ik_{d}}{6}Q_{xz}
- \frac{ik_{d}}{2v_{d}}M_{yz}
- \frac{k_{d}^{2}}{6}O_{xzz}
\right),
\end{align}
\end{subequations}
where $E_{0}$ is the amplitude of the incident electric field, $S_{L}=P^{2}$ denotes the area of the lattice unit cell with periodicity $P$, and $\epsilon_{d}$ is the permittivity of the surrounding medium. Here, $k_{d}$ and $v_{d}$ represent the wave number and phase velocity in the dielectric environment, respectively. The reflectance and transmittance are then obtained as $R = |r|^{2}$ and $T = |t|^{2}$, where the reflection coefficient $r$ is defined analogously (see Ref.~\cite{terekhov2019multipole}).

\subsection{Radiative enhancement factor $\gamma/\gamma_{\infty}$}\label{sec10}
Spontaneous emission is a purely quantum mechanical process. However, the dipole's spontaneous emission rates relative to a reference system are known to be the same under both classical and quantum treatments \cite{xu1999finite}. In classical electromagnetic calculations, the dipole emitter is treated as a radiating point dipole. In order to calculate the radiative decay rate, the total power radiated by the dipole is integrated over a closed surface enclosing the dipole source. The radiative enhancement factor is calculated as $\gamma/\gamma_{\infty} = P/P_{\infty}$ \cite{xu1999finite}, where $P_{\infty}$ is the dipole radiated power that corresponds to the dipole emission in the reference system, that is, \rev{bulk Si}.\\
In case of a metasurface, decay rate enhancement is evaluated through the following simulation procedure: both the silicon (Si) metasurface and the reference system (bulk Si) embedded with a dipole are modeled using a unit cell with periodic boundary conditions applied along the $x$ and $y$ directions to replicate an infinite periodic array. Due to PBC, the dipole in each unit cell is repeated and interacts with the nearby dipole in the periodic structure. To overcome this effect in decay rate calculation, the total emitted power of the dipole in the case of the metasurface  $P$ is scaled by the corresponding total power from the dipole in the reference system $P_{\infty}$. The ratio $ P/P_{\infty}$ quantifies the enhancement of the decay rate.

\revred{To distinguish radiative and non-radiative channels, we define the total, radiative, and non-radiative decay-rate enhancements as}
\begin{equation}
\revred{\frac{\gamma}{\gamma_\infty}=\frac{P}{P_{\mathrm{bulk}}},\qquad
\frac{\gamma_{\mathrm{rad}}}{\gamma_\infty}=\frac{P_{\mathrm{rad}}}{P_{\mathrm{bulk}}},\qquad
\frac{\gamma_{\mathrm{nrad}}}{\gamma_\infty}=\frac{P-P_{\mathrm{rad}}}{P_{\mathrm{bulk}}},}
\end{equation}
\revred{where $P$ is the total emitted power of the dipole in the metasurface environment, $P_{\mathrm{rad}}$ is the radiated power extracted from the outward Poynting flux through the available radiation channels, and $P_{\mathrm{bulk}}$ is the emitted power of the same dipole in homogeneous bulk Si. The radiative efficiency is then evaluated as}
\begin{equation}
\revred{\eta_{\mathrm{rad}}=\frac{P_{\mathrm{rad}}}{P}=\frac{\gamma_{\mathrm{rad}}}{\gamma}.}
\end{equation}
\revred{For the periodic metasurface calculations reported here, $P_{\mathrm{rad}}$ was obtained by integrating the outward time-averaged Poynting flux at the interfaces between the physical domain and the top and bottom perfectly matched layers, whereas $P$ was obtained from the total emitted power of the dipole in the same computational model.}

\section{\label{sec:level4} \textit{CONCLUSIONS} }

\rev{In this work, we employed multipolar decomposition and phase-resolved analysis to investigate symmetry-broken dielectric metasurfaces for spontaneous-emission control.} Using Si cuboid and disk-based metasurfaces as representative platforms, we demonstrated that introducing controlled geometrical asymmetry transforms symmetry-protected dark resonances \cite{liu2018light} into radiative hybrid modes through coherent multipolar coupling. By systematically analyzing the wavelength-dependent multipolar response and its phase evolution, we identified coherent interaction between MD and MQ modes as the dominant mechanism governing the LDOS.
 For Erbium (Er$^{3+}$) ions in broken symmetry Si cuboid and disk-based metasurfaces, this coherent coupling results in strong electromagnetic field confinement at the emitter position, leading to two orders of magnitude spontaneous emission enhancement at 1540 nm, the telecom-C-band. The consistent behavior observed across both cuboid and disk-based metasurfaces underscores the generality and robustness of the proposed mechanism. \rev{The discussion added in this manuscript places this behavior in the broader context of symmetry-broken dielectric resonances, including quasi-BIC and, more generally, symmetry-reduced metasurfaces that can also support chiroptical responses. In contrast to planar chiral metasurfaces, however, the central result here is that symmetry breaking activates a phase-coherent magnetic multipolar channel that enhances the LDOS and controls spontaneous emission from embedded emitters.}
 \revred{Fabrication-tolerance analysis further shows that moderate variations in the notch dimensions and cut angle preserve the resonant enhancement mechanism but mainly lead to spectral detuning, demonstrating that the proposed designs retain practical relevance while still requiring controlled fabrication for exact spectral alignment. This persistence of the resonant response under moderate geometrical deviations indicates that the proposed metasurfaces are not merely idealized structures, but physically meaningful designs in which the underlying MD--MQ hybridization mechanism can survive realistic fabrication irregularities.}
% \revred{ Fabrication-tolerance analysis further shows that moderate variations in the notch dimensions and cut angle preserve the resonant enhancement mechanism but mainly lead to spectral detuning, demonstrating that the proposed designs retain practical relevance while still requiring controlled fabrication for exact spectral alignment.} 
\revred{Beyond the specific metasurface geometries investigated here, the present results are relevant to the development of future optical and quantum-photonic devices, including integrated quantum-photonic platforms, telecom-band single-photon sources, low-loss metasurface emitters, light-emitting devices, and quantum communication technologies. In particular, the identification of coherent MD--MQ interactions as the dominant design mechanism provides a physically transparent route for engineering the local density of optical states and the radiative properties of emitters in dielectric nanophotonic systems.
Thus, this work provides useful design principles for future integrated photonic and quantum-optical devices.}

\section*{Acknowledgments}
M.A.A. would like to acknowledge University Grant Commission (UGC)for providing financial support in the form of Senior Research Fellowship (SRF). F.A.I. would like to acknowledge the financial support from the Department of Science and Technology (DST), India, for funding under the FIST Program 2020 and Core Research Grant (CRG/2021/001167).
\section*{Conflict of Interests}
The authors declare no conflict of interests.

\section*{Data availabilty}
The data set available up on  request to the authors.

% \appendix

% \section{Appendixes}

\bibliography{My_bibliography}% Produces the bibliography via BibTeX.

\end{document}